\documentclass[a4paper,11pt]{article}
\usepackage{pos}
\usepackage{hyperref}
\usepackage{orcidlink}

\def\cascade{{\sc Cascade}}
\def\pythia{{\sc Pythia}}

\def\mcatnlo{{MCatNLO}}

\input epsf.tex
\def\desepsf(#1 width #2){\epsfxsize=#2 \epsfbox{#1}}

\def\pt{\ensuremath{p_{\rm T}}}
\def\PZ{\ensuremath{Z}}

\newcommand{\dphiZ}{{\ensuremath{\Delta\phi_{\PZ\text{j}}}}}

\newcommand{\Zjet}{{$\PZ+$\text{jet}}}

\newcommand{\alphas}{\ensuremath{\alpha_\mathrm{s}}}

\newcommand{\PBM}{PB}

\newcommand{\dphi}{{\ensuremath{\Delta\phi_{12}}}}
\newcommand{\MCatNLO}{{\sc MadGraph5\_aMC@NLO}}

\newcommand{\ptmax}{\ensuremath{\pt^{\text{leading}}}}

\newcommand{\GeV}{\text{GeV}}

\title{High-$p_T$ Azimuthal Correlations of \Zjet\ and Multi-jet Production}
\author*[a]{S.~Taheri~Monfared\orcidlink{0000-0003-2988-7859}}
\author[a]{ L. I. Estevez Banos \orcidlink{0000-0001-6195-3102}}

\affiliation[a]{Deutsches Elektronen-Synchrotron DESY, Germany}

\emailAdd{taheri@mail.desy.de}
\emailAdd{luis.estevez.banos@desy.de}
 
\abstract{In this study, we present our latest findings regarding azimuthal distributions in vector boson + jets and multi-jet production at the Large Hadron Collider (LHC). These findings result from matching next-to-leading order (NLO) perturbative matrix elements with transverse momentum dependent (TMD) parton branching. We conduct a comprehensive comparative analysis of azimuthal correlations between \PZ\   boson-jet and jet-jet systems in the back-to-back region. These distinct azimuthal correlation patterns can help identify potential factorization-breaking effects in this region. Such effects depend on the different color and spin structures of the final states and their interactions with the initial states.
}

\FullConference{The European Physical Society Conference on High Energy Physics (EPS-HEP2023)\\
 21-25 August 2023\\
Hamburg, Germany\\}

\begin{document}

\begin{flushright}
DESY-23-180\\
\end{flushright}

\maketitle

\section{Introduction}
Experiments conducted at the Large Hadron Collider (LHC) involve precise measurements of azimuthal correlations within the context of  \PZ\ bosons + jets  ~\cite{Aad:2013ysa,Khachatryan:2016crw, Aaboud:2017hbk, Sirunyan:2018cpw}  
and  
multi-jet ~\cite{daCosta:2011ni,Khachatryan:2011zj,Khachatryan:2016hkr,CMS:2017cfb,CMS:2019joc}  
final states.  
This comprehensive understanding of these correlations plays a pivotal role in advancing our knowledge of the Quantum Chromodynamics (QCD) sector within the framework of the Standard Model (SM). Additionally, it contributes to the exploration of potential scenarios Beyond-the-SM (BSM).

Azimuthal correlations between \PZ\ bosons and jets or multi-jets provide a sensitive probe of various QCD aspects. At leading order in the strong coupling \alphas, these correlations exhibit a characteristic back-to-back configuration ($\Delta\phi \simeq \pi$), with deviations indicating higher-order QCD radiation. The $\Delta\phi\simeq \pi$ region mainly involves soft gluon radiation, while smaller $\Delta\phi$ regions are associated with hard QCD radiation. The large-$\Delta\phi$ region, where the probed outgoing products are nearly back-to-back, is influenced by both perturbative and non-perturbative QCD contributions, with the balance dependent on the transverse momentum imbalance between the studied products.
The resummation of soft multi-gluon emissions in this region allows the probing of initial state parton transverse momenta, described by transverse momentum-dependent (TMD) parton distribution functions (PDFs) \cite{Angeles-Martinez:2015sea}. Recent theoretical predictions for \PZ boson + jet production, including soft gluon resummation, have been presented \cite{Bouaziz:2022vp,Chien:2022wiq,Buonocore:2021akg}.

In Ref.~\cite{Abdulhamid:2021xtt}, we explored the \dphi\ correlation in high-\pt\ multi-jet events, employing TMD PDFs, parton showers, and NLO calculations. We also highlight the relevance of multi-jet events in studying azimuthal correlations, as they provide sensitivity to factorization-breaking effects~\cite{Collins:2007nk,Vogelsang:2007jk,Rogers:2010dm}, particularly in the presence of colored final states. We propose a systematic comparison between the azimuthal correlation \dphi\ in multi-jet and \Zjet\ production, extending the investigation of azimuthal correlations in \Zjet\ events, as discussed in Ref.~\cite{Chien:2022wiq}, which addresses factorization-breaking.

In Ref.~\cite{Yang:2022qgk}, we offered a detailed comparison of high-\pt\ multi-jet and \Zjet\ production using the \PBM\ TMD method~\cite{Hautmann:2017xtx,Hautmann:2017fcj,Jung:2021mox} matched with NLO calculations. Our multi-jet study ~\cite{Abdulhamid:2021xtt} demonstrated that NLO \PBM\ TMD predictions accurately describe multi-jet azimuthal correlations~\cite{Sirunyan:2017jnl,Sirunyan:2019rpc}. In Ref.~\cite{Yang:2022qgk}, we extend the same method to calculate \Zjet\ production and present corresponding predictions. By employing the same kinematic region for high-\pt\ multi-jet and \Zjet\ production, we enable a direct comparison of angular observables between the two cases.

\section{NLO matching with PB TMD}

The PB approach~\cite{Hautmann:2017xtx,Hautmann:2017fcj,Jung:2021mox} provides 
evolution equations for TMD distributions in terms of Sudakov form factors 
and splitting probabilities, and a corresponding TMD parton shower in a backward 
evolution scheme.  
PB TMD distributions and parton showers are implemented in the 
Monte Carlo event generator {\sc Cascade}3~\cite{Baranov:2021uol}. 

A method to match  TMD evolution with NLO perturbative matrix elements 
has been developed for the case of the Drell-Yan process in 
Refs.~\cite{BermudezMartinez:2019anj,BermudezMartinez:2020tys} 
using the framework of {\sc MadGraph5\_aMC@NLO}~\cite{Alwall:2014hca}. 
We next apply this method to the case 
of the jet production process~\cite{Abdulhamid:2021xtt}, matching 
 PB TMD distributions and parton showers 
with  multi-jet NLO matrix elements from 
 {\sc MadGraph5\_aMC@NLO}. 
Further details on the NLO matching method with PB TMD are given in 
Ref.~\cite{Yang:2022qgk},  
where a comparison of 
  MCatNLO+{\sc Cascade}3 \cite{Baranov:2021uol} 
and 
MCatNLO+{\sc Herwig}6 \cite{Corcella:2002jc} 
matching 
is performed. 

\section{Azimuthal correlations in  \Zjet\  and multi-jet production}
\label{sec:correlations}

\begin{figure}[h!tb]
\begin{center} 
\includegraphics[width=0.49\textwidth]{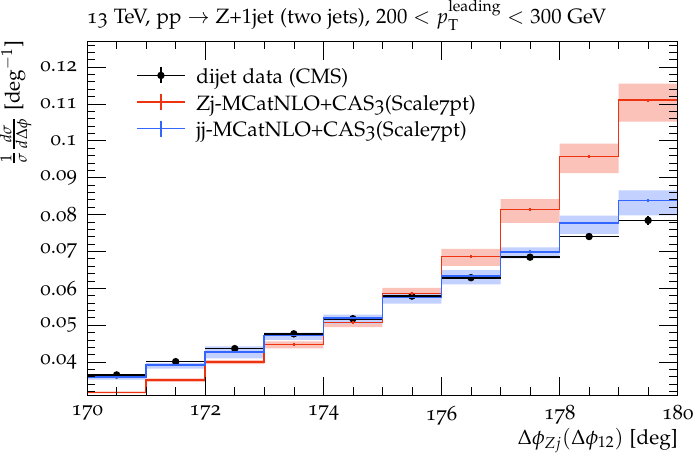} 
\includegraphics[width=0.49\textwidth]{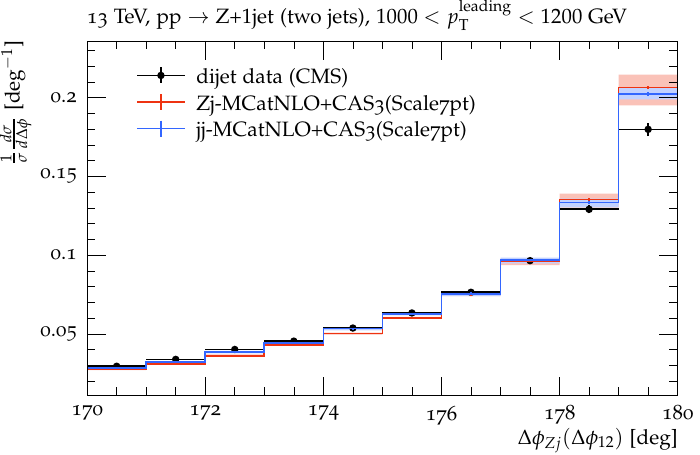} 
  \caption{\small Azimuthal correlation predictions (\dphiZ\ and \dphi) for \Zjet\ and multi-jet processes are shown in the back-to-back region, both for $\ptmax > 200 $ \GeV\ (left) and $\ptmax > 1000 $ \GeV\ (right). These predictions use the \mcatnlo+CAS3 framework, with depicted uncertainties from scale variation. We include CMS multi-jet correlation measurements~\cite{Sirunyan:2019rpc} for reference.
}
\label{b2b-Zets_CAS}
\end{center}
\end{figure}

In Fig.\ref{b2b-Zets_CAS}, we present the results of our matching method. Specifically, we display the predicted azimuthal correlations \dphiZ\ for \Zjet\ production in the back-to-back region. To provide context, we also include predictions for azimuthal correlations \dphi\ in multi-jet production within the same kinematic region. These predictions are compared to multi-jet production measurements obtained by CMS~\cite{Sirunyan:2019rpc}.
 Our findings reveal that the \dphiZ\ distribution in \Zjet\ production, for $\ptmax >200$ \GeV, exhibits a stronger correlation towards $\pi$ in comparison to the \dphi\ distribution in multi-jet production. This discrepancy diminishes for $\ptmax > 1000$ \GeV.

Variations in $\Delta \phi$ between \Zjet\ and multi-jet production arise from differences in the initial state's flavor composition, transverse momenta, and parton showers. These distinctions also result from variations in final state showering due to the differing numbers of colored final state partons. Factorization-breaking effects, caused by interference between initial and final state partons, depend on the final state's structure and the number of colored final state partons.

Another key point regarding Fig.\ref{b2b-Zets_CAS} is the strong agreement between multi-jet production measurements and \mcatnlo+CAS3 predictions, as extensively discussed in Ref.\cite{Abdulhamid:2021xtt}. Notably, this agreement holds for most cases, except at very high \ptmax, where a deviation from measurements appears. This discrepancy might indicate a factorization violation. Thus, it is crucial to measure $\Delta\phi$ distributions in other factorization-consistent processes.

Our findings reveal that, in the region of leading transverse momenta around $\pt \approx {\cal O}(100)$ GeV, the boson-jet final state exhibits stronger azimuthal correlations than the jet-jet final state. As transverse momenta increase beyond the electroweak symmetry breaking scale ($\pt \approx {\cal O}(1000)$ GeV), this difference diminishes, and the boson-jet and jet-jet correlations become more similar. We attribute this behavior to the characteristics of partonic initial state and final state radiation in both cases. As potential factorization-breaking effects arise from color interferences of initial-state and final-state radiation, different breaking patterns are expected for strong and weak azimuthal correlations, impacting the boson-jet and jet-jet cases differently. Consequently, we advocate for a systematic comparison of multi-jet and \Zjet\ distributions, spanning the phase space from low transverse momenta ($\pt \approx {\cal O}(100)$ GeV) to high transverse momenta ($\pt \approx {\cal O}(1000)$ GeV).

\begin{figure}[h!tb]
\begin{center} 
\includegraphics[width=0.45\textwidth]{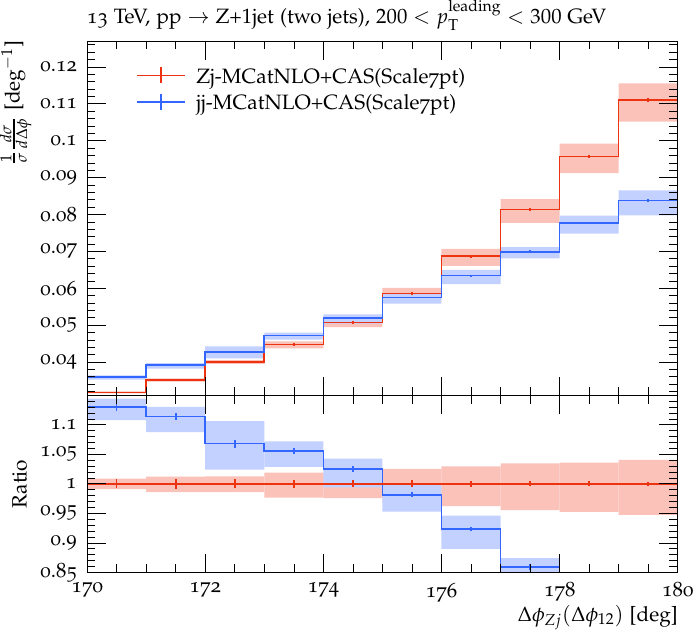} 
\includegraphics[width=0.45\textwidth]{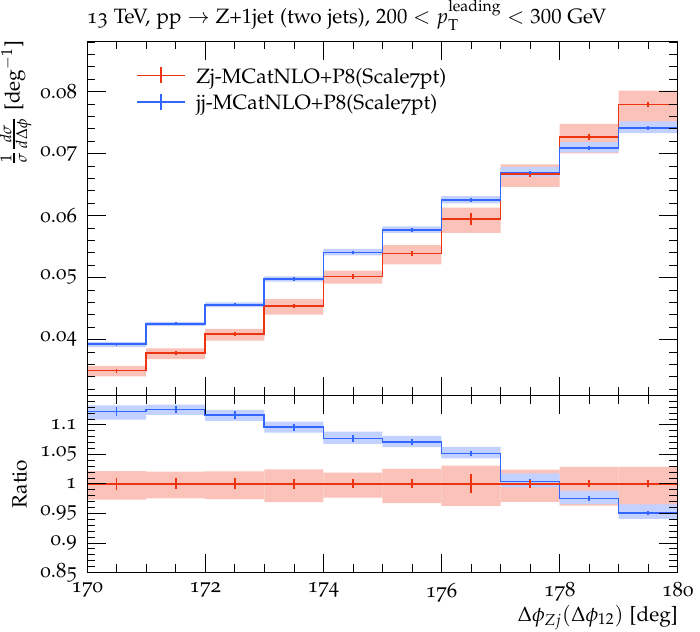} 
\includegraphics[width=0.45\textwidth]{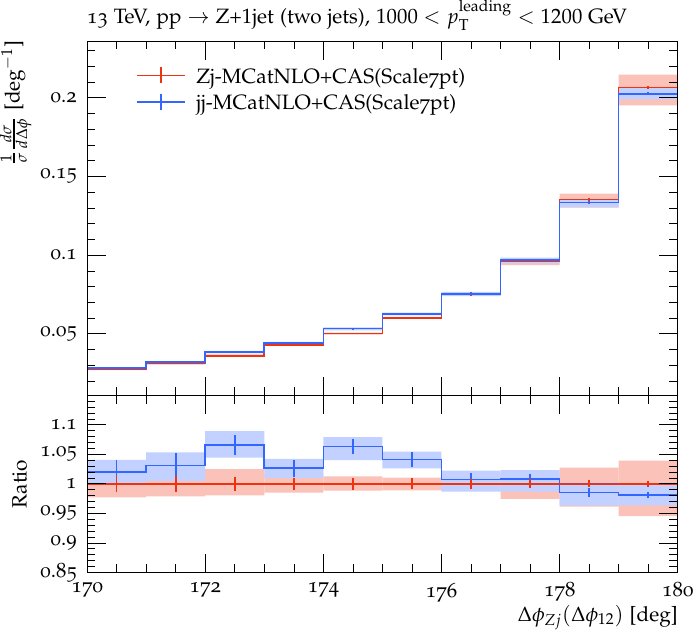} 
\includegraphics[width=0.45\textwidth]{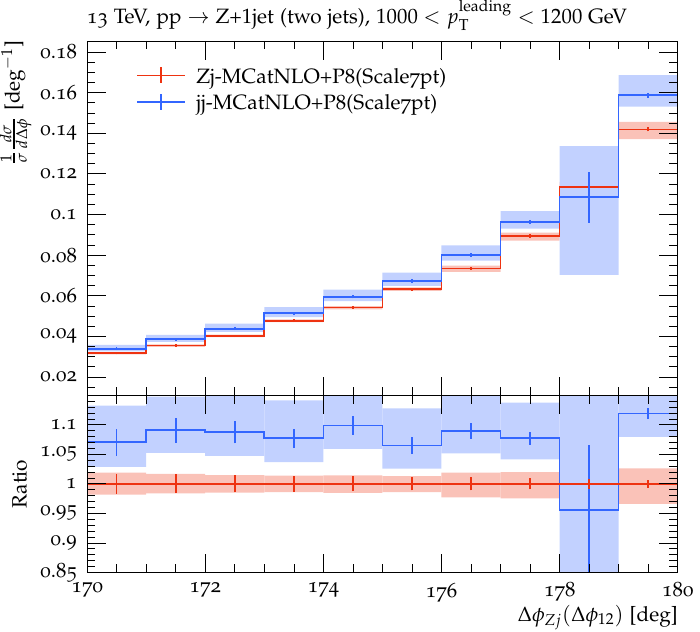} 
  \caption{\small Predictions for the azimuthal correlation \dphiZ (\dphi)  in the back-to-back region for \protect\Zjet\ and multi-jet production obtained with \protect\mcatnlo+CAS3 (left column) and  \protect\mcatnlo+\pythia 8 (right column). Shown are different regions in  $\ptmax > 200 $~\GeV\ (upper row) and $\ptmax > 1000 $~\GeV\ (lower row). The bands show the uncertainties obtained from scale variation.  }
\label{b2b-Zets_CAS_P8}
\end{center}
\end{figure}

Assessing the impact of final state radiation on the \dphi\ correlations proves challenging, as the subtraction terms in the NLO matrix element computation are also contingent on the final state parton shower characteristics. To gauge the influence of final state showering, we compare \dphi\ correlation calculations in the back-to-back region achieved through two methods: \mcatnlo+CAS3 and \mcatnlo+\pythia 8 (as displayed in Fig.\ref{b2b-Zets_CAS_P8}). In the \mcatnlo+\pythia 8 calculation, we incorporate \pythia 8's subtraction terms into the \MCatNLO\ framework, utilize NNPDF3.0 parton density, and tune CUETP8M1~\cite{Khachatryan:2015pea}.

Fig.~\ref{b2b-Zets_CAS_P8} illustrates discrepancies in the distributions stemming from the distinct parton showers employed in \cascade\ and \pythia 8. Nevertheless, it's noteworthy that the ratio of distributions between \Zjet\ and multi-jet production follows a consistent trend: for \Zjet -production, the distribution is steeper (indicating stronger correlation) at lower \ptmax, while at higher \ptmax, the distributions adopt similar shapes. From this, we deduce that the primary contributor to $\Delta\phi$ decorrelation is initial state radiation. Consequently, the shape of the $\Delta\phi$ decorrelation in the back-to-back region aligns between \Zjet\ and multi-jet processes at high \ptmax\ when similar initial partonic states come into play.

\section{Conclusion}
\label{sec-conc}

In this article  we have discussed predictions from PB TMD parton showers  for  final state observables at the LHC,   focusing on the azimuthal correlations  of multi-jets and \Zjet. 
 The azimuthal correlations \dphiZ , obtained in \Zjet\  production are steeper compared to those in jet-jet production (\dphi ) at transverse momenta ${\cal O}(100)$ \GeV , while they become similar for very high transverse momenta, ${\cal O}(1000)$ \GeV , which is a result of  similar initial parton configuration of both processes.

In \Zjet\ production the color and spin structure of the partonic final state is different compared to the one in multi-jet production, and differences in the azimuthal correlation patterns can be used to search for  potential 
{factorization - breaking} effects, involving initial and final state interferences. In order to experimentally investigate those effects, we propose to measure the ratio of the distributions in \dphiZ \ for \Zjet\  and \dphi\ for multi-jet production at low and at very high \ptmax , and compare the measurements to predictions obtained assuming that factorization holds.

\section*{Acknowledgments}
The findings presented in this paper draw upon the research detailed in Ref. \cite{Abdulhamid:2021xtt,Yang:2022qgk}. Many thanks to all co-authors for collaboration. We are grateful to the organizers of EPS2023 for the invitation to present these results at the workshop.

\end{document}